   \documentclass[a4paper,12pt]{article}
\usepackage{times,mathptm,amsmath,amssymb}

\usepackage{graphicx,here,cite,fancyhdr}
\usepackage[margin=15mm]{geometry}
\begin{document}


\vspace*{13mm}
\centerline{\bf\LARGE What is novel in quantum transport for mesoscopics?
\footnote{Invited talk presented at the 50th Golden Jubilee DAE
Solid State Physics Symposium, BARC, Mumbai, 2005.}}

\vspace*{8mm}

\centerline{
{\bf\large Mukunda P. Das}$^{\dagger}$
{\bf\large and Frederick Green}$^{+}$
}

\centerline{
$^{\dagger}${Department of Theoretical Physics, IAS,
The Australian National University,
Canberra, ACT 0200, Australia}}
\centerline{
$^{+}${School of Physics, The University of New South Wales,
Sydney, NSW 2052, Australia}}


\begin{abstract}
{The understanding of mesoscopic transport has now attained an
ultimate simplicity. Indeed, orthodox quantum kinetics would seem to
say little about mesoscopics that has not been revealed -- nearly
effortlessly -- by more popular means. Such is far from the case,
however. The fact that kinetic theory remains very much in charge is
best appreciated through the physics of a quantum point contact. While
discretization of its conductance is viewed as the exclusive result
of coherent, single-electron-wave transmission, this does not begin
to address the paramount feature of all metallic conduction: dissipation.
A perfect quantum point contact still has finite resistance, so its
ballistic carriers must dissipate the energy gained from the applied
field. How do they manage that? The key is in standard many-body
quantum theory, and its conservation principles.}
\end{abstract}



\parindent4mm

\noindent
{\bf {INTRODUCTION}}

\noindent
A striking signature of mesoscopic transport, as evidenced in quantum
point contacts (QPCs), is the discretization of conductance into
``Landauer steps'' in units of $2e^2/h$. The steps appear to be well
described by the coherent transmission of independent electron waves
through the contact, imagined as a perfectly lossless quantum barrier
\cite{p1,p2,p3}.

The techniques of single-particle scattering are universally accessible.
Consequently, the pioneering insights of the Landauer school
have achieved more than to open a new vista of small-scale device physics.
They have also made their compact understanding available to one and all,
in a toolbox of easily grasped phenomenological design aids.

Besides the great simplicity of this widely adopted approach,
it is generally agreed that it cannot sustain any contradiction with
the older-established principles of microscopic transport theory
\cite{mahan}.
In this overview we clarify, in somewhat sharper detail than usual,
those interrelationships that may exist between canonical kinetics
on the one hand, and Landauer transport theory on the other.

Central to the Landauer description are two assumptions:
(a) current flows in a QPC when a {\em mismatch} of chemical potentials
is set up across the ends of the wire. In response, carriers flow
more or less freely from the ``high-density'' lead end to the
``low-density'' lead end. (b) The intervening channel is a quantum
tunnelling barrier that moderates the unimpeded flux of electrons.
The Landauer conductance formula directly encodes this tunnelling physics.

Assumptions (a) and (b) are not consistent with each other.
The first asserts the validity of charge drift in the mesoscopic
regime: a difference in electron density across the leads drives a
current from the high to the low region. That is, the current flow
is {\em metallic} because it engages carrier states that are
well filled and spatially extended. In contradistinction, (b) asserts
instead that the flux rate is set -- quite literally -- by the
probability of tunnelling through some phenomenologically chosen barrier.
(Surprisingly, if such a barrier had any internal physical structure
it would be irrelevant to the outcome.) Here the physics is that
of states separately confined to the leads on either side of the barrier.
All that matters is that they have some {\em residual overlap}.

The relation between the two hypotheses itself invites two questions.
To what extent do we have a picture -- case (a) -- of metallic conduction
involving truly extended states, and to what extent is it a description
-- case (b) -- of (self-evidently non-metallic) tunnelling involving
spatially separate, autonomous in- and out-states? Should one conclude,
astonishingly, that mesoscopic transport is really both at the same time?

These are intriguing, if incidental, issues. We now discuss the
other thought-provoking aspects of the relation between Landauer
phenomenology and conventional quantum kinetics.

\noindent
{\bf {THE PHYSICAL PROBLEM: DISSIPATION}}

\noindent
The essential role (or, more accurately, the essential
absence) of {\em resistive energy dissipation} has recently returned
to the forefront of discussions about the microscopic basis
of the Landauer approach
\cite{davies,agrait,vpt}.
In a perceptive early critique of Landauer's picture
(wherein ballistic conduction consists in perfectly coherent and,
therefore, exclusively elastic transmission), Frensley
\cite{frensley}
had already identified the singular lack of a theoretical account
of dissipation (Joule heating) within the Landauer phenomenology.
In admitting coherent scattering, and that only, as the origin of
QPC conductance, the Landauer model leaves an enormous unhealed
gap between it and the fluctuation-dissipation theorem
\cite{kubo},
which universally quantifies the conductance in terms of the actual
energy loss via the dissipative electron-hole pair processes
that always accompany metallic transport
\cite{mahan}.

In their more recent discussions, Davies
\cite{davies}
and Agra\"{\i}t {\em et al}
\cite{agrait}
have also covered the unresolved status of ballistic dissipation.
We also have summarised our own considerations from the standpoint
of kinetics
\cite{vpt}.
To date the problem seems to have had no deeper analysis on the part
of any proponents of the Landauer philosophy, other than that
dissipation occurs somehow, somewhere, deep in the leads and far
from the active channel
\cite{p2};
too far off, anyway, to spoil the undisputed
simplicity of the coherent-tunnelling account of ballistic conductance.

Surveys of the dissipation issue all agree
\cite{davies,agrait,vpt,frensley}
that na\"{\i}ve quantum mechanical descriptions of single-carrier
tunnelling are unable to settle the central problem of conduction:
{\em What causes the dissipation in a ballistic QPC}?
The matter goes well beyond this simple academic consideration.

Not too long from now, reliable and effective nano-electronic
design will grow to demand, not models that are built for minimum
effort, but ones that are microscopically grounded and therefore
credible, both as basic physics and as quantitatively trustworthy
engineering tools. Device designers, above all, will need every confidence
to predict the dominant {\em dissipative} characteristics of their
new quasi-molecular structures, operating far away
from equilibrium. The Landauer approach is not made for such demands.

We now review the answer to the question posed. Its resolution
has been available all the while -- definitively, and free of any
artificial conundrums -- within many-body quantum kinetics
\cite{mahan,DG1}.
The microscopic application of many-body methods leads not only
to conductance quantization by fully accounting for inelastic energy loss
\cite{DG2},
but it also resolves a long-standing experimental enigma
\cite{rez}
in the noise spectrum of a quantum point contact (QPC)
\cite{DG3}.
Evidently, the same developments will foreshadow a
systematic pathway to the truly predictive design of
novel structures.

\noindent
{\bf {QUANTUM KINETICS: THE SOLUTION}}

\noindent
The central issue in conduction is clear.
Any finite conductance $G$ must dissipate electrical energy at the rate
$P = IV = GV^2$, where $I = GV$ is the current and $V$ the potential
difference across the terminals of the driven conductor.
Physically, there must be an explicit mechanism
(e.g. emission of optical phonons) through which the energy gained
by carriers, when driven from source to drain, is channelled to
the surroundings.

Alongside all the elastic and coherent scattering processes,
inelastic processes must also act. Connected with elastic
single-particle scattering, intimately and inevitably, are
its dynamic and dissipative companions:
the electron-hole vertex corrections
\cite{mahan}.
This much is required by the conservation laws
for the electron gas itself
\cite{pinoz}.
Over and above this (and still within the global purview of
conservation), there will be additional decay modes
coupling the electrons to other background excitations.

None of these dissipative effects can be described at the level
of simple, one-particle coherent quantum mechanics, for they are
all {\em inherently many-body} effects, requiring a genuinely
microscopic description.
Harnessed together, the elastic and inelastic processes fix $G$.
Yet it is only the energy-dissipating mechanisms that secure
the thermodynamic stability vital to steady-state conduction.

We already possess a complete quantum-kinetic understanding
of the ubiquitous power-loss formula $P = GV^2$
\cite{kubo,sols,wims}.
It resides in the fluctuation-dissipation theorem,
valid for {\em all} resistive devices at all scales,
without exception. The theorem expresses
the requirement for thermodynamic stability.
A plain chain of reasoning follows from it:
\cite{DG1,DG2}

(i) inelasticity is necessary and sufficient to stabilize current
flow at finite conductance;

\vspace{-4mm}
(ii) ballistic quantum point contacts have finite $G \propto 2e^2/h$;
therefore

\vspace{-4mm}
(iii) the physics of energy loss is indispensable
to a proper theory of ballistic transport.

\noindent
The physics of {\em explicit} inelastic scattering is beyond the scope 
of transport models that rely only on coherent quantum scattering
to explain the origin of $G$ in quantum point contacts. Coherence
implies elasticity, and elastic scattering is always loss-free:
it conserves the energy of the scattered particle. This reveals
the deficiency of purely elastic models of transmission. We now
review a well-defined microscopic remedy for this deficiency.

To allow for the energy dissipation vital to any microscopic
description of ballistic transport, we recall that open-boundary
conditions imply the intimate coupling of the QPC channel to its interfaces
with the reservoirs. The interface regions must be treated
as an integral part of the device model. They are the very sites
for strong scattering effects: dissipative many-body events
as the current  enters and leaves the ballistic channel,
and elastic one-body events as the carriers interact
with background impurities, the potential barriers that confine and funnel
the current, and so on. 

The key idea in our standard treatment is to subsume the
interfaces within the total kinetic description of the ballistic channel.
At the same time, strict charge conservation in an open
device requires the direct supply and removal of current
by an external generator
\cite{sols}
or, equally well, a battery
\cite{wims};
that is, a source for the driving field that itself
is outside the system.
These two criteria are equivalent. They are also prescriptive;
mandated by electrodynamics whenever a metallic channel
is subjected to {\em external} electromotive forces.

In truly open-system operation, therefore, the current
is determined externally, independent of the local physics
peculiar to the reservoirs.
This canonical requirement sets the quantum kinetic approach entirely
apart from the Landauer multi-reservoir scenario
\cite{p1},
which rests upon a purely intuitive phenomenology: that the
current has to depend on hypothetical density gradients
between the reservoirs.
For an externally driven charged system, orthodox electrodynamics
{\em never} entails this.

\noindent
{\bf QPC CONDUCTANCE: ITS QUANTUM KINETIC EMERGENCE}

\noindent
It is straightforward to write the algebra for the
ideal ``Landauer'' conductance in our model system. A uniform,
one-dimensional ballistic QPC, of operational length $L$,
will be associated with two mean free paths
determined by $v_{\rm F}$, the Fermi velocity of the electrons,
and a pair of characteristic scattering times
$\tau_{\rm el},\tau_{\rm in}$. Thus

%
\vspace{-6mm}
\begin{equation}
\lambda_{\rm el} = v_{\rm F}\tau_{\rm el};{~~}
\lambda_{\rm in} = v_{\rm F}\tau_{\rm in}.
{\bf \label{e1}}
\end{equation}
\vspace{-6mm}

\noindent
Respectively, these are the scattering lengths set by the
elastic and inelastic processes active at both interfaces. The
device (i.e. the QPC {\em with} its interfaces) has a conductive
core that is strictly collisionless. It follows from this ballistic
boundary condition that $L$ then delimits both elastic and
inelastic mean free paths, leading directly to

%
\vspace{-6mm}
\begin{equation}
\lambda_{\rm el} = L = \lambda_{\rm in}.
{\bf \label{e2}}
\end{equation}
\vspace{-6mm}

Finally, the channel's conductance is given by the familiar formula

%
\vspace{-6mm}
\begin{equation}
G = {ne^2\tau_{\rm tot}\over m^*L}
= {2k_{\rm F}\over \pi}{e^2\over m^*L}
{\left( {\tau_{\rm in}\tau_{\rm el} \over
         {\tau_{\rm el} + \tau_{\rm in}}} \right)};
{\bf \label{e3}}
\end{equation}
\vspace{-6mm}

\noindent
the effective mass of the carriers is $m^*$.
In the first factor of the rightmost expression for $G$
we rewrite the density $n$ in terms of the Fermi momentum $k_{\rm F}$;
in the final factor, we use Matthiessen's rule
$\tau_{\rm tot}^{-1} = \tau_{\rm el}^{-1} + \tau_{\rm in}^{-1}$
for the total scattering rate in the system.

\noindent
Using Equations (\ref{e1})--(\ref{e3}), the conductance
reduces to

%
\vspace{-6mm}
\begin{equation}
G = 2{e^2\over \pi\hbar}{\hbar k_{\rm F}\over m^* L}
{\left( {(L/v_{\rm F})^2 \over 2L/v_{\rm F}} \right)}
= {2e^2\over h} \equiv G_0.
{\bf \label{e4}}
\vspace{-6mm}
\end{equation}

\noindent
This is precisely the Landauer conductance of a
single, one-dimensional, ideal channel.

None of the conjectural assumptions,
otherwise invoked to explain conductance quantization
\cite{p1,p3},
has been used. Rather, it is the axioms of electrodynamics
and microscopic response theory, {\em and only those},
which guarantee Eq. (\ref{e4}). In fact, we have just seen
directly how this distinctive mesoscopic result emerges from
completely standard quantum kinetics.

\begin{figure}[H]
 \centering
 \includegraphics[width=.45\textwidth]{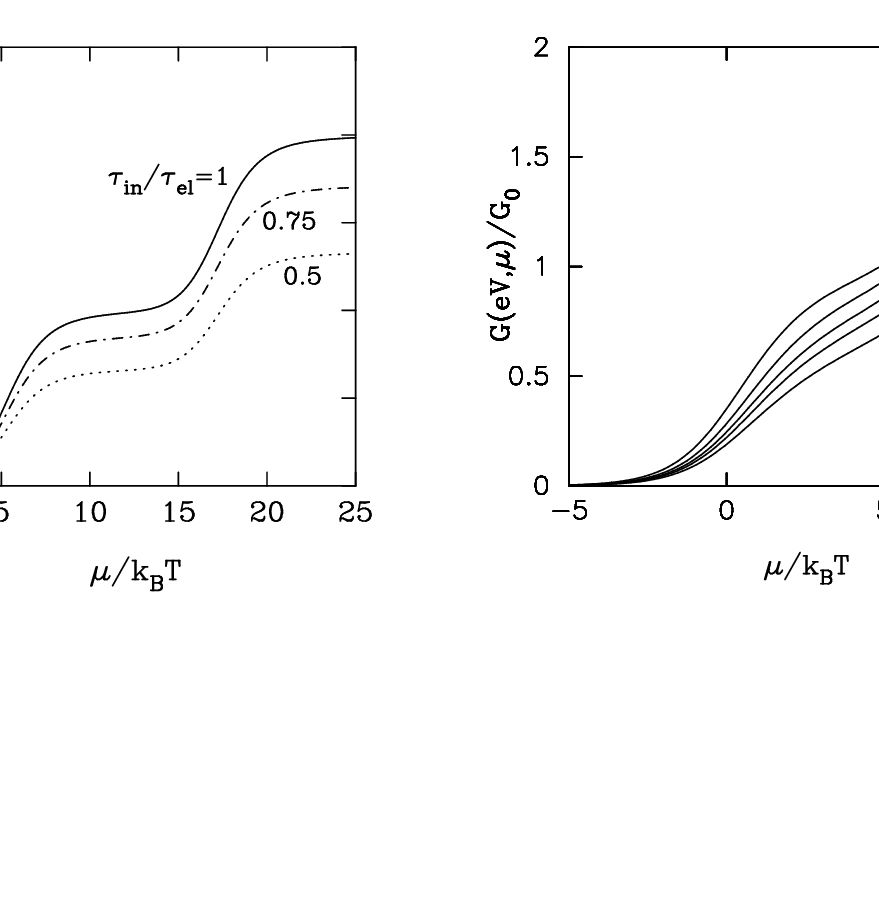}
\vspace{-30mm}
\caption{\small quantization of conductance in a two-band ballistic
quantum point contact, calculated within our kinetic theory
(see Ref. \cite{DG1}) and displayed as a function of
chemical potential $\mu$ in thermal units. The plots are
normalised to units of the Landauer conductance quantum $G_0$.
{\em Left panel}, full curve: $G$ for an idealised ballistic channel.
Broken curves: non-ideal behaviour increases
with the onset of inelastic phonon emission inside the contact.
{\em Right panel}: non-equilibrium conductance for an actual
multi-band QPC, computed using a realistic model for field-dependent
inelastic scattering. The device characteristics correspond to
those of a heterojunction-based channel fabricated on epitaxial
GaAs/AlGaAs. Compare Fig. 2 of Reznikov {\em et al}.
\cite{rez}.}
\end{figure}
\vspace{-4mm}

In the left-hand panel of Figure 1 we plot the results of our
model for a QPC
\cite{DG1}
consisting of two one-dimensional conduction bands with their
threshold energies separated by 12$k_{\rm B}T$, in thermal units
at temperature $T$. We use the natural extension of Eq. (\ref{e4})
to cases where one or more channels may be open to conduction,
depending on $T$ as well as the size of the chemical potential $\mu$.
As the role of inelastic scattering is enhanced
($\tau_{\rm in} < \tau_{\rm el}$) the conductance deviates
from the ideally ballistic ``Landauer'' limit.

The right-hand panel of Fig. 1 shows, as a precursor to our
discussion of non-equilibrium noise (see also Fig. 3 below),
the same conserving quantum-kinetic calculation
of $G$ for a realistic point contact.
It is noteworthy that the non-ideality of $G$
in the quasi-one-dimensional quantum channel grows progressively
as the source-drain voltage that drives the mesoscopic current
increases: thus the overall value of $G$ goes down as the voltage
runs from low (0.5mV) to high (3mV), and the rate of optical-phonon
emission increases for carriers accelerated within the channel.
For a comparison with corresponding results, as extracted
from raw measurements made in a QPC sample, see Fig. 2 of Ref.
\cite{rez}.

Most important to the microscopic derivation of the Landauer
quantized conductance is the clear and central role of inelastic
energy loss, one of the underpinnings of quantum transport.
Charge conservation, the other underpinning,
is guaranteed by our use of microscopically
consistent open-boundary conditions at the interfaces.
Their importance cannot be emphasised too strongly.
It is fair to say that, as essential physical requirements,
they are not transparent within some of the more intuitive
derivations of Eq. (\ref{e4}).

\noindent
{\bf QPC NOISE: REVEALING MICROSCOPICS}

\noindent
The noise response of a quantum point contact is a fascinating
aspect of mesoscopic transport, and a more demanding one both
experimentally and theoretically. In 1995, a landmark measurement
of non-equilibrium noise was performed by the Weizmann group
\cite{rez},
which yielded a very puzzling result. Whereas conventional models
\cite{p3}
fail to predict any structure {\em at all} in their noise signal
for a QPC driven at constant current levels,
the data show an orderly series of
marked and increasingly strong peaks, just where the carrier density
in the QPC starts to grow and becomes metallic.

Remarkable as they are to this day, the Weizmann results remained
absolutely unexplained for a decade.
We have now accounted for the Reznikov {\em et al.} measurements,
within our strictly conserving kinetic description
\cite{DG3}.

In the following Fig. 2 we display the experimental data
side by side with our
computation of excess QPC noise, under the same conditions
\cite{DG3}.
In contrast to the outcome of popular mesoscopic phenomenology
\cite{p3}
one notes the close affinity between the measurements and the quantum
kinetic calculation,
as the carrier density is swept across the first conduction-band
threshold, where the conductance exhibits its lowest step.
At fixed values of source-drain current, the accepted noise models
predict no peaks at all, but rather a featureless monotonic drop
in the noise strength as the carrier density passes through threshold.
\vspace{2ex}\

\begin{figure}[H]
 \centering
 \includegraphics[width=0.85\textwidth]{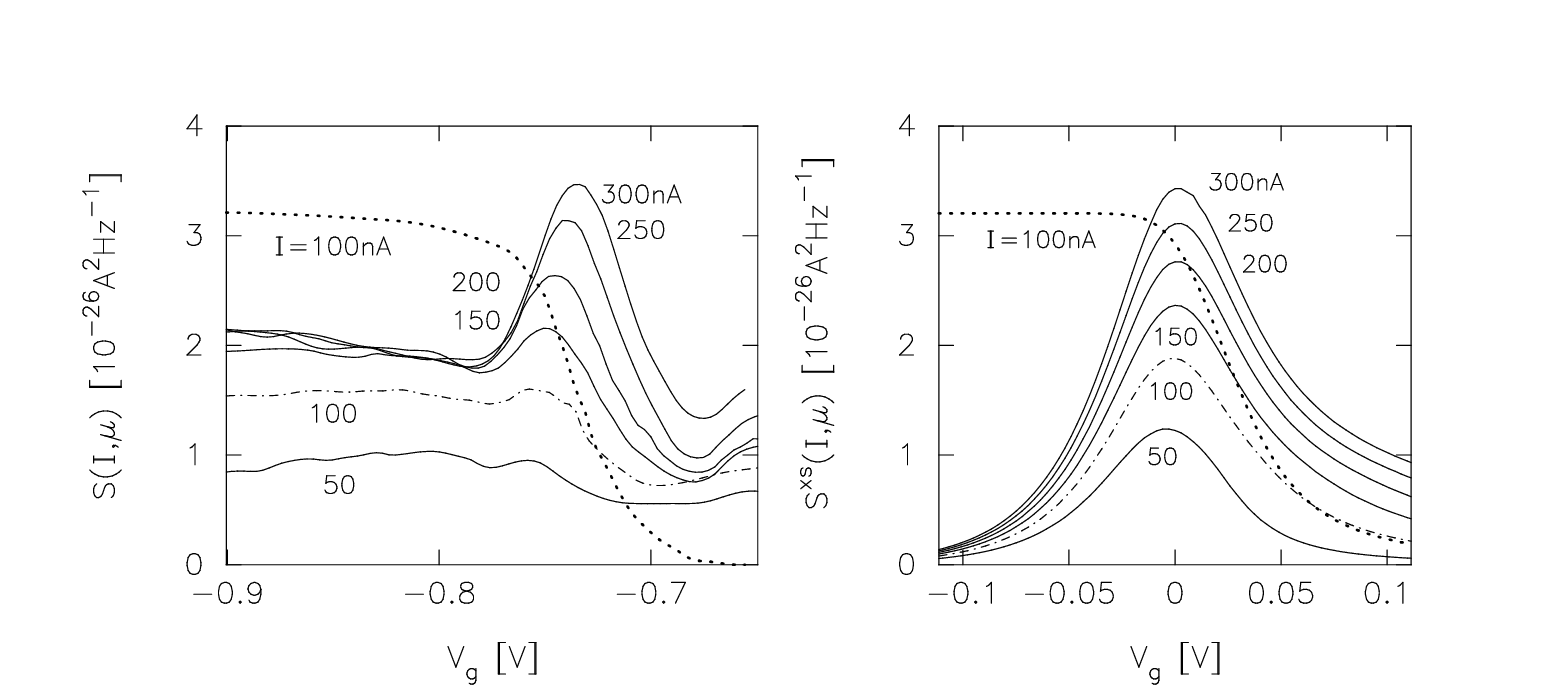}
\vspace{-5mm}
 \caption{\small Non-equilibrium current noise of a QPC at constant
source-drain current, as a function of gate bias. Left: data
from Reznikov {\em et al}, Ref. \cite{rez}. Right: microscopically
conserving kinetic calculation from Green {\em et al.}, Ref. \cite{DG3}.
In each case the
dotted line traces the currently accepted shot-noise prediction at 100nA
using, as respective inputs, measured and calculated data for $G$.
The prediction from Landauer theory is well wide of the mark.}
\end{figure}

To round off our survey of ballistic noise we present
a final figure. It serves to show that the remarkable peaks observed
in a QPC channel, at constant applied current, are by no means
fortuitous artifacts. In Fig. 3 (the QPC noise measured concomitantly
with $G$ in the right-hand-side panel of Fig. 1) we see an
unfolding, characteristic peak sequence at
{\em constant source-drain voltage} as the gate bias
systematically pushes the conduction electrons upward in density:
first though the lowest, and then the next higher, sub-bands
in the structure. The noise maxima are very well replicated by
the physics built into our conserving microscopic description.
Once again they invite favourable comparison with observations.
This can be checked against Fig. 2 of Ref.
\cite{rez}.

The constant-voltage peaks have been widely celebrated
as the predictive triumph of mesoscopic transport phenomenology
\cite{p3}.
On that score, any alternative fluctuation theory for QPCs
must do at least as well. What is unique about
our quantum kinetic approach is not that it offers a
microscopically founded account for effects already explained
in the Landauer framework. What is really different is that it
describes, faithfully, everything else that phenomenology has
signally failed to predict in the noise spectrum at constant current.

\begin{figure}[H]
 \centering
 \includegraphics[width=0.40\textwidth]{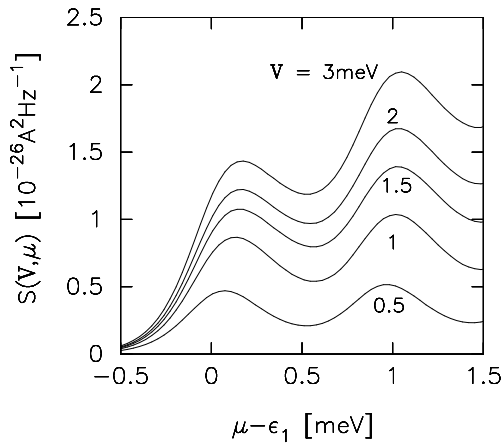}
\vspace{-24mm}
 \caption{\small Non-equilibrium noise in a QPC at constant
levels of voltage, computed quantum kinetically as for Fig.2
\cite{DG3}.
Characteristic peak sequences appear at each of the
two lowest sub-band energy thresholds. Note also the presence
of a slower-rising signal background on which the relatively
sharper maxima are superimposed. A similar background is
evident in the corresonding experimental data
\cite{rez}.
While such maxima at constant voltage are predicted by other
approaches
\cite{p3},
the rising background -- like the unexpected
structures at constant current -- are not reproduced except
by the fully kinetic model.}
\end{figure}


If there is a single, utterly fundamental, reason why orthodox
quantum kinetics is able to open up in such a striking way
the microscopics of fine-scale fluctuations, it is that --
unlike other explanations of transport and noise -- it starts
its exploration from the universal principles of conservation,
and also ends with them intact.
This is exactly the reason for kinetic theory's freedom from
any and all unjustified, {\em ad hoc} assumptions. It is time
to look at the mesoscopic action of conservation.

\noindent
{\bf CENTRALITY OF CONSERVATION}

\noindent
The key to all quantum kinetic descriptions of conductance is the
fluctuation-dissipation theorem, whose practical implementation is
Eq. (\ref{e3}) (where the overall relaxation time $\tau_{\rm tot}$
encodes all the electron-fluctuation dynamics via the Kubo formula
\cite{kubo}). This universal relation is one of the electron-gas
{\em sum rules}
\cite{pinoz}.
In this instance, it expresses the conservation of energy, dissipatively
transferred from an external source to the thermal surroundings, for
any process that involves resistive transport -- including
that in a ballistic quantum point contact.

A second, and equally fundamental, sum rule concerns the compressibility
of an electron fluid in a conductive channel. This sum rule turns out
to have an intimate link with the non-equilibrium noise behavior reviewed above;
here we give a brief explanation of that crucial link. For details see Refs.
\cite{DG3} and \cite{csr}-\cite{bbk}.

Recall that the carriers in a quantum point contact are stabilized
by the presence of the large leads, which pin the electron density
to fixed values on the outer boundaries of the interfaces (recall too
that the interfaces and the channel together define the open system).
No matter what the transport processes within the QPC may be,
or how extreme, the system's {\em global neutrality} is guaranteed
by the stability of the large and charge-neutral reservoirs.

It follows that the total number $N$ of active electrons in the
device remains independent of any current that is forced
through the channel, for $N$ is always neutralized by the ionic
background in its neighborhood, as well as the stabilizing leads.
The presence of the latter means that all remnant fringing fields
are screened out beyond the device boundaries;
hence the global neutrality.

One can then prove that the total mean-square number fluctuation
$\Delta N = k_{\rm B}T{\partial N/\partial \mu}$
is likewise independent of the external applied current
\cite{csr}.
The compressibility of the carriers in the QPC is given
in terms of $N$ and $\Delta N$ by
\cite{pinoz}

\vspace{-6mm}
\begin{equation}
\kappa = {L\over Nk_{\rm B}T} {\Delta N\over N},
{\bf \label{e5}}
\end{equation}
\vspace{-6mm}

\noindent
which, in consequence, remains strictly unaffected by any transport process.
This a surprising corollary of global neutrality. It asserts that, in
an open conductor, the system's equilibrium compressibility completely
determines the compressibility of the electrons when driven away from
equilibrium, {\em regardless} of how strong the driving field is.

The compressibility sum rule expresses the unconditional
conservation of carriers in a non-equilibrium conductor.
Previously unexamined in mesoscopics, this principle has
an immediate importance and applicability. 

How does $\kappa$ determine the noise in a QPC? The strength
of the current fluctuations is, at base, the product of two
contending factors:

\vspace{-6mm}
\begin{equation}
S(I,t) \sim {\langle I(t)I(0) \rangle}
{\Delta N\over N}.
{\bf \label{e6}}
\end{equation}
\vspace{-6mm}

\noindent
The first factor represents the self-correlation of the
instantaneous electron current $I(t)$ evaluated as a trace
over the non-equilibrium distribution of excited
electrons in the device. The second factor -- evidently
a basic characteristic of the electron gas in the channel -- is
independent of $I$, meaning that the invariant compressibility
(Eq. (\ref{e5})) must dictate the overall scale of the noise spectrum.

Let us examine the noise spectra of Figs. 2 and 3
in light of this key result
\cite{DG3}.

\vspace{-4mm}
\begin{itemize}
\item
At large negative bias $V_g$, the channel is depleted. The remnant
carriers are classical, so $\Delta N/N \to 1$. The noise is then
dominated by strong inelastic processes at high driving fields,
as embodied in ${\langle I(t)I(0) \rangle}$.
\item
In the opposite bias limit (right-hand sector of each panel
in Fig. 2), the channel is richly populated and thus
highly degenerate, with a large Fermi energy $E_{\rm F}$.
Then $\Delta N/N \to k_{\rm B}T/2E_{\rm F} \ll 1$. The noise
spectrum falls off according to Eq. (\ref{e6}), since the
current-correlation factor -- now well within the regime of
ballistic operation -- reaches a fixed ideal value.
\item
In the mid-range of bias voltage, there is a point where the
carriers' chemical potential matches the energy threshold for
populating the first conduction sub-band. Here there is a robust
competition: on the one hand, scattering processes that reduce
the correlation ${\langle I(t)I(0) \rangle}$ are less effective,
while on the other hand the onset of degeneracy drives the compressibility
down. Where this interplay is strongest, there are peaks.
\end{itemize}
\vspace{-4mm}

Now we understand the outcome of the compressibility rule: it is,
quite directly, the ``inexplicable'' emergence of the noise peak structures.
The striking case of QPC noise gives an insight into the
central importance of the conserving sum rules in the physics
of transport at meso- and nanoscopic dimensions.
The more imaginative treatments of noise fail to address the
explicit action of microscopic conservation in ballistic phenomena,
and therefore cannot offer a rational understanding
of the real nature of ballistic conduction. 

\noindent
{\bf {SUMMARY}}

\noindent
In this paper we have recalled the most fundamental aspects of
mesoscopic transport physics, and the need to make sure that
descriptions of it continue to respect those aspects. They are:
the {\em primacy of microscopic conservation} in charged open systems; the
{\em dominance of many-body phenomena}, most of all dissipation;
and the {\em unity of conductance and fluctuations}. 

Quantum kinetic theory was, and remains, the sole analytical
method that can guarantee all of these requirements.
It provides a detailed, cohesive and inherently microscopic
account of conductance and noise together. This applies to
the specific case of quantum point contacts.
In our own quantum-kinetic studies we have accurately
reproduced -- free of any and all special pleadings -- the proper current
response of a mesoscopic channel, including the quantized-conductance
signature. 

Given the keys to this standard and yet newly fruitful picture
(open-system charge conservation and the efficacy
of dissipative many-body scattering), our quantum kinetic
analysis presents as a thoroughly orthodox development.
Precisely because it is so firmly and conventionally grounded,
it affords an unambiguous, natural and quantitative
understanding of the non-equilibrium fluctuations of
a quantum point contact, with its associated dynamics.
That understanding has been successfully tested in fully explaining
the long-standing puzzle posed by the noise measurements
of Reznikov {\em et al.}
\cite{rez}.

The theoretical impact of noise and fluctuation physics is that it
carries much more information on the internal dynamics of mesocopic
systems -- a level of knowledge inaccessible through the $I$-$V$
characteristics on their own.
The capacity for a self-contained microscopic explanation
of mesoscopic transport processes underwrites a matching
ability to build new programs for device design that are 
inherently rational. The {\em practical} need for such programs
can hardly be overstated in the context of nanotechnology,
given the ongoing paucity of physically based techniques for it.

In the authors' view, the time is ripe to restore the methods
of microscopic quantum kinetics to a central place in
mesoscopic electronics, where they seem to have been much less
in evidence over recent years. The need is manifest, and is
becoming more and more pressing with the advance of technology.
Some of the basic tools to meet it are already at hand
\cite{DG2,kubo,sols,wims,pinoz,csr,bbk}.

%
\noindent
{\bf {REFERENCES}}
\vspace{-20mm}

\renewcommand{\refname}{}



\begin{thebibliography}{0}
\setlength{\itemsep}{-1ex}

\bibitem{p1} Y. Imry, {\em Introduction to Mesoscopic Physics},
2nd Edition (Oxford University Press, Oxford, 2002)

\bibitem{p2}
Y. Imry and R. Landauer, Rev. Mod. Phys. {\bf 71}, S306 (1999)

\bibitem{p3} Y. M. Blanter and M. B\"uttiker,
Phys. Rep. {\bf 336}, 1 (2000)

\bibitem{mahan}
G. D. Mahan, {\em Many-Particle Physics} 3rd Edition
(Plenum, New York, 1990) Ch 7

\bibitem{davies}
J. Davies, {\em Physics of Low-Dimensional Semiconductors:
An Introduction} (Cambridge University Press, Cambridge, 1998)
p 199 ff

\bibitem{agrait}
N. Agra\"{\i}t, A. Levy Yeyati, and J. M. van Ruitenbeek,
Phys. Rep. {\bf 377} 81 (2003); see Section  IIID5

\bibitem{vpt}
M. P. Das and F. Green, J. Phys.: Condens. Matter
{\bf 17}, V13 (2005)

\bibitem{frensley}
W. R. Frensley, in {\em Heterostructures and Quantum Devices}
Eds. W. R. Frensley and N. G. Einspruch (Academic, San Diego, 1994) Ch 9

\bibitem{kubo}
R. Kubo, M. Toda, and M. Hashitsume,
{\it Statistical Physics II: Non-equilibrium Statistical Mechanics},
2nd Edition (Springer, Berlin, 1991)

\bibitem{DG1}
M. P. Das and F. Green,
J. Phys.: Condens. Matter {\bf 15}, L687 (2003)

\bibitem{DG2}
F. Green and M. P. Das,
J. Phys.: Condens. Matter {\bf 12}, 5233 (2000); {\em ibid}, 5251

\bibitem{rez}
M. Reznikov, M. Heiblum, H. Shtrikman, and D. Mahalu,
Phys. Rev. Lett. {\bf 75}, 3340 (1995)

\bibitem{DG3}
F. Green, J. S. Thakur, and M. P. Das,
Phys. Rev. Lett. {\bf 92}, 156804 (2004)

\bibitem{pinoz}
D. Pines and P. Nozi\`eres,
{\em The Theory of Quantum Liquids} (Benjamin, New York, 1966)

\bibitem{sols}
F. Sols, Phys. Rev. Lett. {\bf 67}, 2874 (1991)

\bibitem{wims}
W. Magnus and W. Schoenmaker, {\em Quantum Transport in
Sub-micron Devices: A Theoretical Introduction} (Springer, Berlin, 2002)

\bibitem{csr}
J. S. Thakur, F. Green, and M. P. Das,
Int. J. Mod. Phys. {\bf B18}, 1479 (2004)

\bibitem{ppt}
M. P. Das, J. S. Thakur, and F. Green,
ArXiv preprint cond-mat/0401134 (2004)

\bibitem{bbk}
F. Green and M. P. Das in {\em Noise and Fluctuations Control in
Electronic Devices}, A. A. Balandin ed (American Scientific Publishers,
Stevenson Ranch, 2002), pp 31--48


\end{thebibliography}
\end{document}